\documentclass[referee]{aa}
\usepackage{epsfig,deluxe}
\begin{document}

\newcommand{\gsim}{\hbox{\rlap{$^>$}$_\sim$}}
  \thesaurus{06;  19.63.1}

\authorrunning{S. Dado, A. Dar \& A. De R\'ujula}
\titlerunning{Afterglow of GRB 011121}
\title{SN Search in the Afterglow of GRB 011121}

\author{Shlomo Dado$^{^1}$, Arnon Dar$^{^1}$ and
A. De R\'ujula$^{^2}$}
\institute{1. Physics Department and Space Research Institute, Technion\\
               Haifa 32000, Israel\\
           2. Theory Division, CERN, CH-1211 Geneva 23, Switzerland}

\maketitle

\begin{abstract} 

In order to assist in planing the search of supernova (SN) in the late time
afterglow of the gamma ray burst (GRB) 011121 we use the Cannonball Model
of GRBs and the early time observations of its afterglow (AG) in the R 
band to predict its late time decline and intrinsic spectrum. 

\end{abstract}

On Nov. 21, 18:47:21 UT a very bright GRB (011121) has been simultaneously
detected and localized by BeppoSAX (Piro 2001a,b) and IPN (Hurley et al.
2001).  Its optical afterglow was first detected 10.3 hours after the
burst in the R-band (Krzysztof et al. 2001).  Further observations
reported 4 additional R band values for the fading source during the first
two days after burst (Staneck et al. 2001a,b,c) and a possible redshift $\rm
z=0.36$ (Infante et al. 2001b) for the host galaxy. A possible candidate
for this host galaxy was detected 0.5'' (approximately 5 kpc) from the GRB
location (Stanek et al. 2001a). As was pointed out by Stanek et al.
(2001b), the relatively low redshift and fast decay of its afterglow (AG)
make GRB 011121 an attractive search-target for a possible supernova (SN)
associated with this burst.  In order to assist in planing this search we
have used the Cannonball Model (CB) of GRBs (Dado, Dar and De R\'ujula
2001 and references therein) and these early time R-band observations,
recalibrated by the photometry of Olsen et al. (2001), to predict the late
time behaviour of the AG in the BVRI bands. The predictions are shown in
Figs.  1-5. They are explained shortly below. Detailed derivations can be
found in Dado et al. (2001). 

In the CB model, long-duration GRBs and their AGs are produced in core
collapse supernovae by jets of highly relativistic ``cannonballs'' that
pierce through the supernova shell. The AG --the persistent radiation in
the direction of an observed GRB-- has three origins: the ejected CBs, the
concomitant SN explosion, and the host galaxy. These components are
usually unresolved in the measured ``GRB afterglows'', so that the
corresponding light curves and spectra are the cumulative energy flux
density: 

\begin{equation}
\rm    F_{AG}=F_{CBs}+F_{SN}+F_{HG}\, ,
\label{sum}
\end{equation}                      

\noindent 
The contribution of the candidate host galaxy depends on the
angular aperture of the observations and it is usually determined 
by late time observations when the CB and SN contributions become 
negligible.

\noindent
Core-collapse supernovae (SNII/Ib/Ic)
are far from being standard candles. But if their explosions
are fairly asymmetric ---as they would be if a fair fraction of
them emit two opposite jets of CBs---  much of the variability could be a 
reflection of the varying angles from which we see their
non-spherically expanding shells. 
Exploiting this possibility to its extreme, we shall use
SN1998bw as an ansatz standard candle (Dado et al. 2001 and references 
therein):
Let the energy flux density of SN1998bw 
at redshift $\rm z_{bw}=0.085$ (Galama et al. 1998)
be $\rm F_{bw}[\nu,t]$.
For a similar SN placed at a redshift $\rm z$:
\begin{eqnarray}
{\rm F_{SN}[\nu,t] = } &&
{\rm{1+z \over 1+z_{bw}}\;
{D_L^2(z_{bw})\over D_L^2(z)}}\, \times\nonumber \\
&&{\rm F_{bw}\left[\nu\,{1+z \over 1+z_{bw}},t\,     
{1+z_{bw} \over 1+z}\right]\; A(\nu,z)}\, ,
\label{bw}
\end{eqnarray}
where $\rm D_L(z)$ is the luminosity distance\footnote{The cosmological
parameters we use in our calculations are:
$\rm H_0=65$ km/(s Mpc), ${\rm \Omega_M}=0.3$ and
${\rm \Omega_\Lambda}=0.7$.}
and $\rm A(\nu,z)$ is the extinction along
the line of sight. The extinction in our Galaxy in the direction 
of GRB 011121 
is E(B-V)=0.5 (Schlegel et al. 1998) which yields an attenuation 
factor $\rm A(\nu,z)=0.34 $ ($\rm A_R=1.16$, Olsen et al. 2001),
but  for the line of sight in the host galaxy it will have to be
estimated from the spectra of its observed AG and host galaxy.

\noindent
The contribution of a CB to the GRB afterglow is given by 
(Dado et al. 2001):                                  
\begin{equation}
\rm
F_{CB}=f \; [\gamma(t)]^{2\alpha}\;[\delta(t)]^{3+\alpha}\,
\nu^{-\alpha} ,
\label{fluxdensity2}
\end{equation} 
where $\rm f$ is a normalization constant (see Dado et al. 2001
for its theoretical estimate), 
$\rm \alpha\approx -1.1$ is 
the spectral index of the synchrotron radiation from Fermi accelerated electrons in the CB whose
acceleration rate is in equilibrium with their cooling rate by
synchrotron emission, $\rm \gamma(t)$ is the Lorentz factor of   
the CB and $\rm \delta(t)$ is its Doppler factor,
\begin{equation}
\rm
\delta\equiv\rm{1\over\gamma\,(1-\beta\cos\theta)}
\simeq\rm {2\,\gamma\over (1+\theta^2\gamma^2)}\; ,
\label{doppler}
\end{equation}
whose approximate expression is valid for small observing angles
$\theta\ll 1$ and $\gamma\gg 1$,
the domain of interest for GRBs.      
For an interstellar medium of constant baryon density $\rm n_p$, the Lorentz 
factor, $\rm\gamma(t)$ is given by (Dado et al. 2001): 
\begin{eqnarray}
\rm \gamma&=&\rm\gamma(\gamma_0,\theta,x_\infty;t)
=\rm {1\over B} \,\left[\theta^2+C\,\theta^4+{1\over C}\right]\nonumber\\
\rm C&\equiv&\rm
\left[{2\over B^2+2\,\theta^6+B\,\sqrt{B^2+4\,\theta^6}}\right]^{1/3}
\nonumber\\
\rm B&\equiv&\rm
{1\over \gamma_0^3}+{3\,\theta^2\over\gamma_0}+
{6\,c\, t\over  (1+z)\, x_\infty}               
\label{cubic}
\end{eqnarray}
where $\gamma_0=\gamma(0)$ and
\begin{equation}
\rm 
x_\infty\equiv{N_{CB}\over\pi\, R_{max}^2\, n_p}
\label{range} 
\end{equation}
is the distance travelled by the CB until it stops,
$\rm N_{CB}$ is its baryon number and $\rm R_{max}$
is its radius.

The best fitted parameters to the observed early time R-band  AG 
of GRB 011121 are, $\alpha = 1.10\, ,$
$\rm \gamma_0 = 1302\, , $ $\rm \theta = 0.084\, mrad\, , $ and $\rm
x_\infty = 1.10\, Mpc\, .$ The resulting late time R-band light curves
are presented in Figs. 1,2. The contributions of the host galaxy 
$\rm F_{HG}\approx 2\, \mu J$ $(\rm R_{host}\approx 23)$ and the SN to
the light curves should be considered as upper bounds since their
true values depend, respectively, on the angular aperture of the
observations and on the unknown extinction in the host galaxy. We also
present in Figs. 3-5 the CB model predictions for the light curves of the
AG in the BVI bands with the host galaxy's contribution subtracted. The SN
contribution to the BVI bands was reduced by the Galactic extinction in
these bands in the direction of GRB 011121. These preliminary predictions
can be refined when more observational data become available.

In conclusion, if GRB 011121 has the relatively low redshift, z=0.36, a SN
akin to 1998bw should be observed around December 10 as can be seen in
Fig. 1, unless its light is absorbed considerably in the host galaxy. The
parameters we have determined in the CB model from the preliminary data
have considerable uncertainties, but if its relatively low redshift is
confirmed, the conclusion is inescapable that the supernova associated
with GRB 011121 will, at about day 20, tower in the BVRI bands above the
proper GRB's afterglow: $\rm F_{SN}\sim 30-100\, F_{CB}$, as can be seen in
Figs. 2-5.

\newpage
\begin{figure}[]
\hskip 2truecm
\vspace*{2cm}
\hspace*{-1.7cm}
\epsfig{file=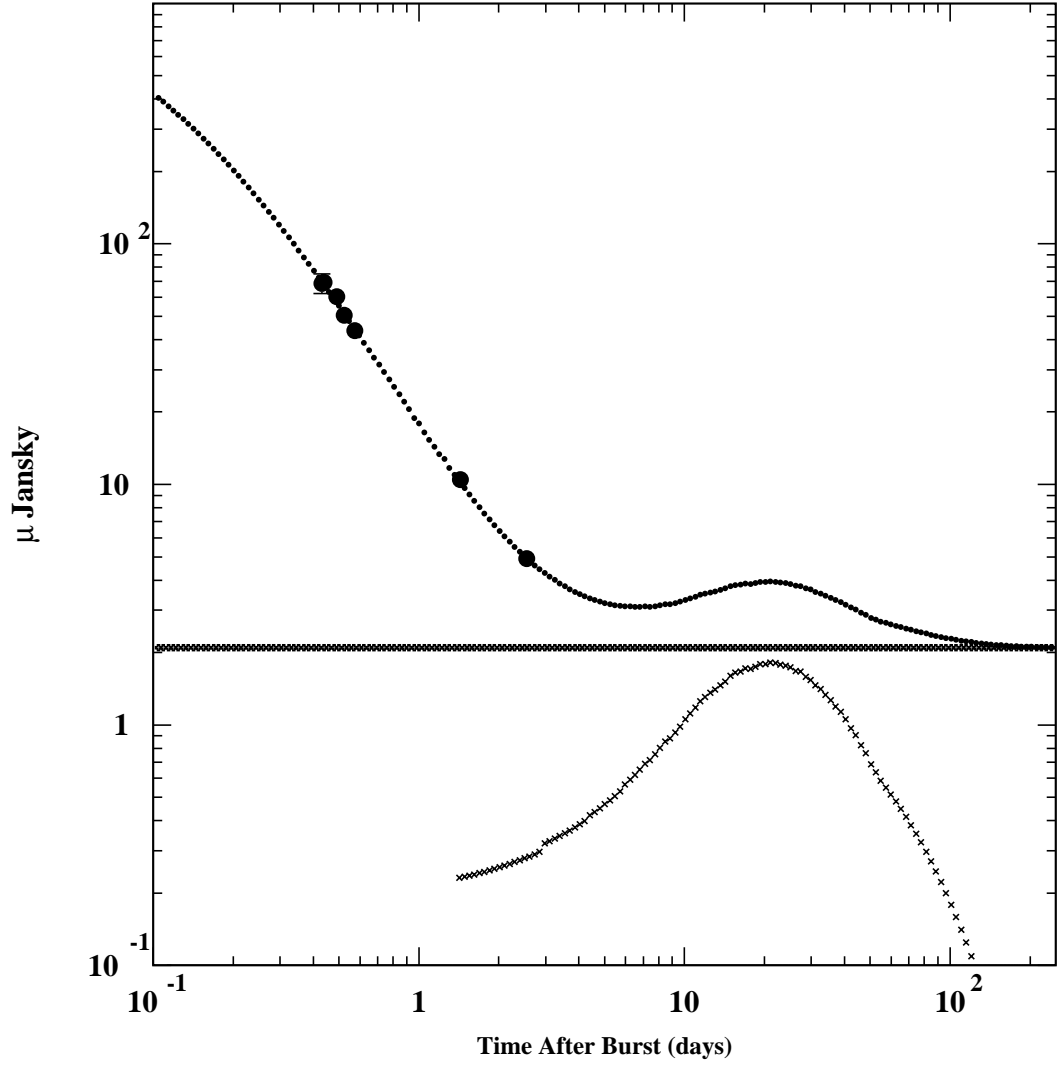, width=16cm} 
\caption{ Comparisons between our fitted R-band afterglow
and the observations not corrected for  extinction
for GRB 011121 at $\rm z=0.36\,, $
without subtraction of the host
galaxy's contribution (the straight line). The
CB's AG (the line of squares) is given by Eqs.~3-5~. 
The contribution from a 1998bw-like supernova placed at the GRB's
redshift and modified by Galactic extinction, Eq.~(\ref{bw}), 
is indicated by a line of crosses.
The SN 1998bw-like contribution, is mildly detectable.}
\label{fig1121}
\end{figure}  
\newpage

%%%%%%%%%%%%%%%%%%%%%%%%%%%%%%%%%%%%%%%%%%%%%%%%%%%%%%%%
\begin{figure}[]
\hskip 2truecm
\vspace*{2cm}
\hspace*{-1.7cm}
\epsfig{file=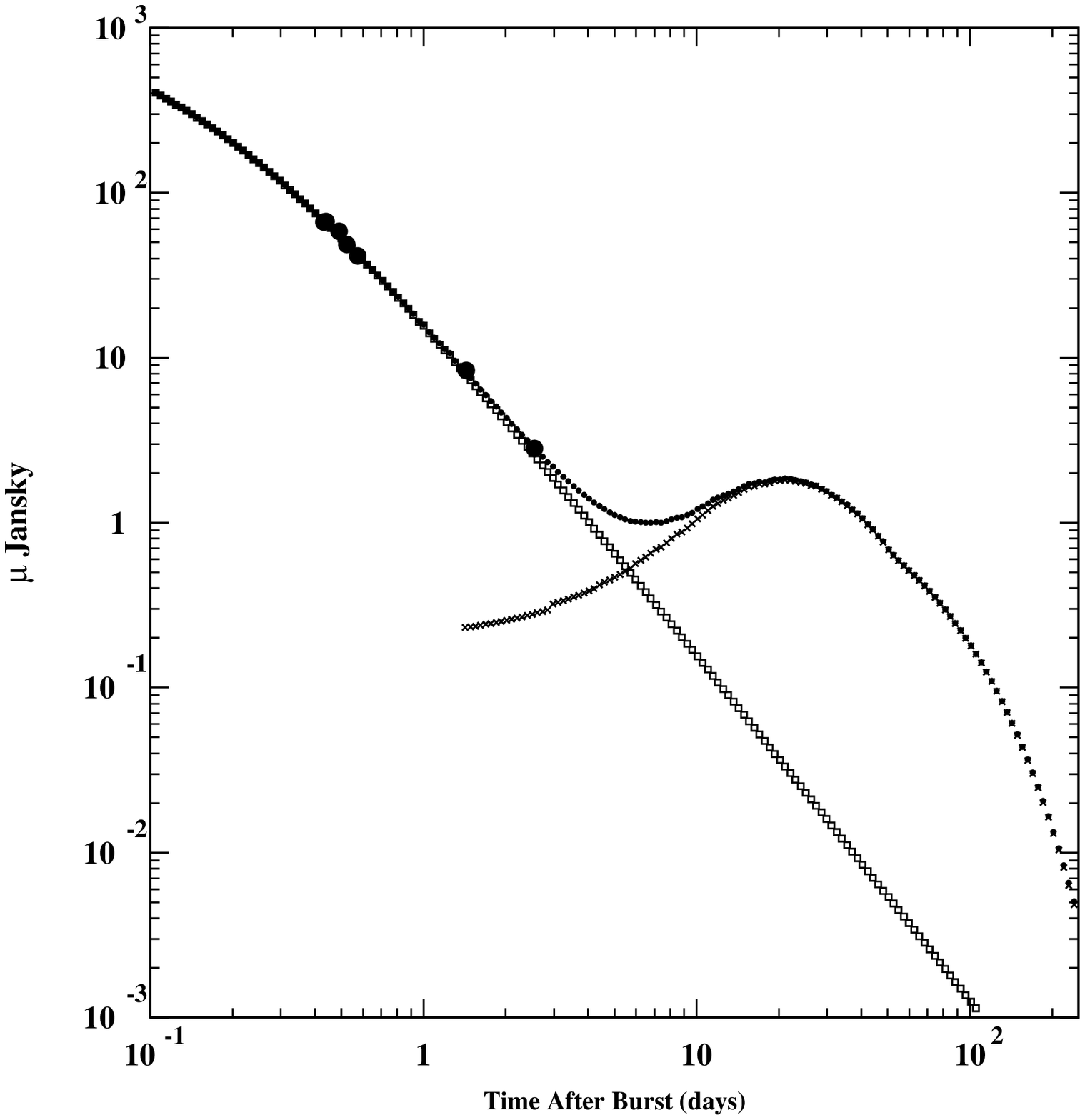, width=16cm} 
\caption{ Comparisons between our fitted R-band afterglow
and the observations 
not corrected for  extinction with the host galaxy's contribution 
subtracted, for GRB 011121 at $\rm z=0.36$. 
The CB's AG (the line of squares) is given by Eqs.~3-5~.  
The contribution
from a 1998bw-like supernova placed at the GRB's
redshift and modified by Galactic extinction, Eq.~(\ref{bw}), 
is indicated by a line of crosses.}
\label{figr1121}
\end{figure}  

%%%%%%%%%%%%%%%%%%%%%%%%%%%%%%%%%%%%%%%%%%%%%%%%%%%%%%%%
\begin{figure}[]
\hskip 2truecm
\vspace*{2cm}
\hspace*{-1.7cm}
\epsfig{file=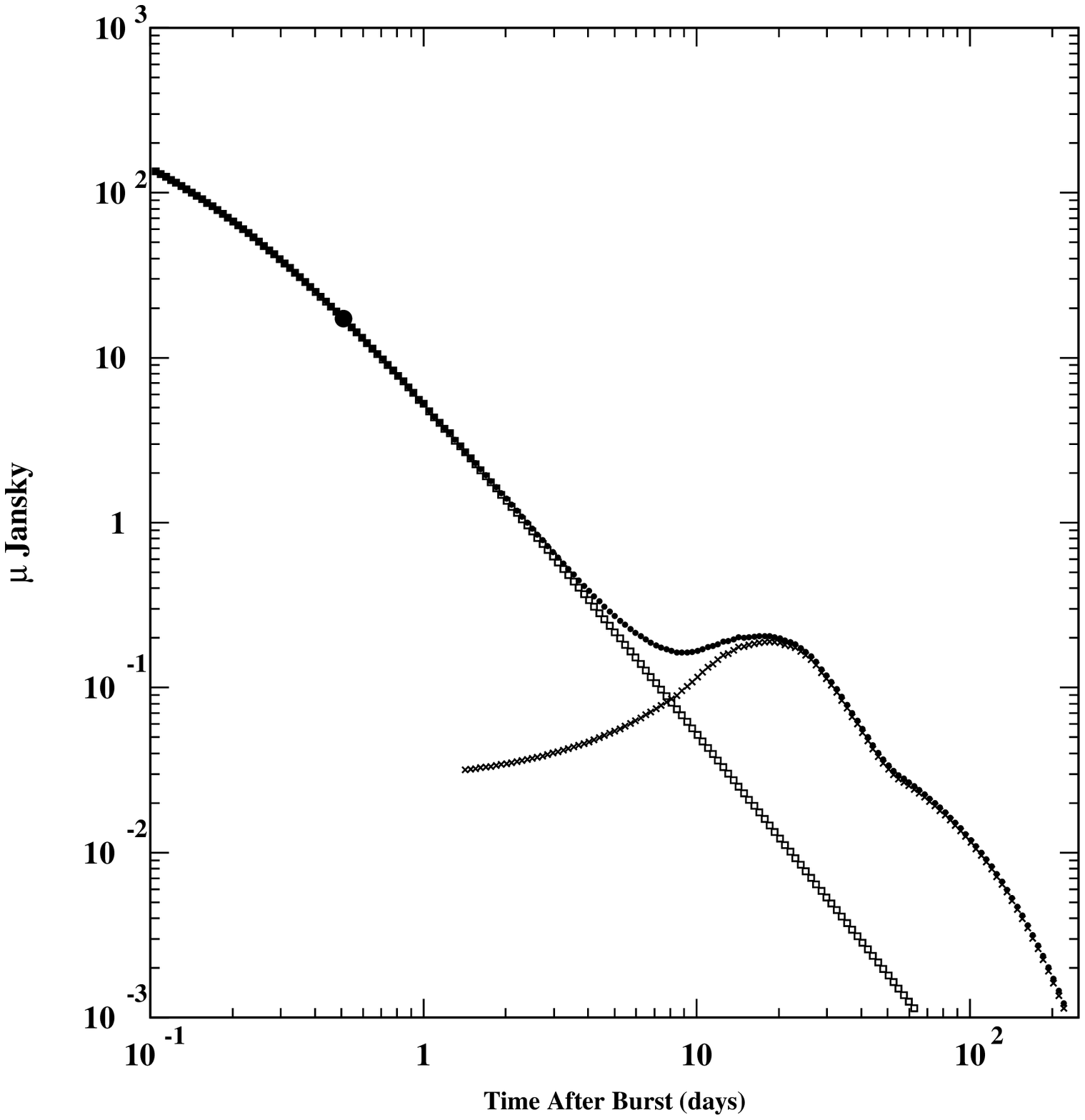, width=16cm} 
\caption{ Comparisons between our fitted B-band afterglow
and the observations 
not corrected for  extinction with the host galaxy's contribution 
subtracted, for GRB 011121 at $\rm z=0.36$. 
TheCB's AG (the line of squares) is given by Eqs.~3-5~.  
The contribution
from a 1998bw-like supernova placed at the GRB's
redshift and modified by Galactic extinction, Eq.~(\ref{bw}),
is indicated by a line of crosses.}
\label{figb1121}
\end{figure}  

%%%%%%%%%%%%%%%%%%%%%%%%%%%%%%%%%%%%%%%%%%%%%%%%%%%%%%%%
\begin{figure}[]
\hskip 2truecm
\vspace*{2cm}
\hspace*{-1.7cm}
\epsfig{file=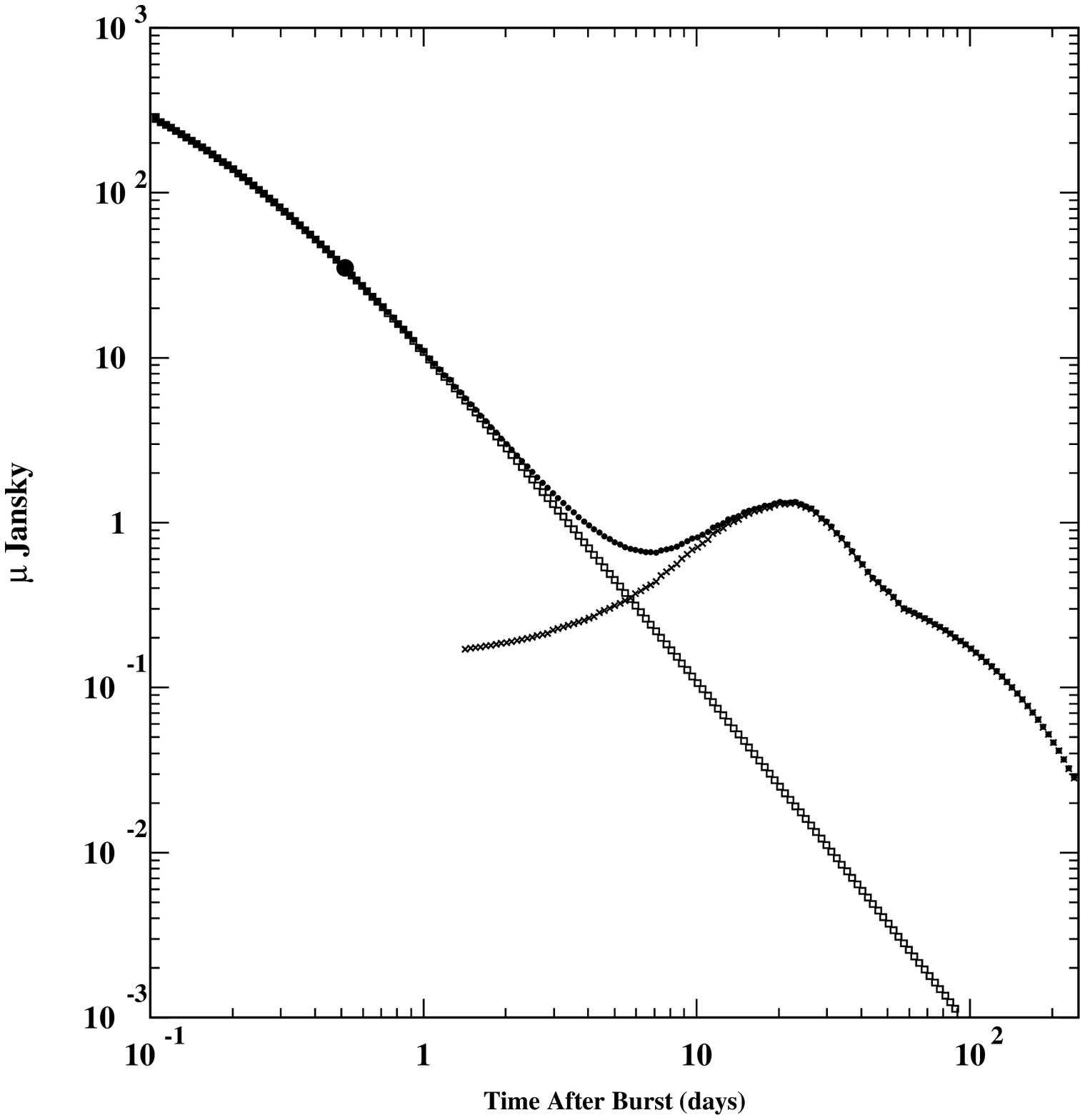, width=16cm} 
\caption{ Comparisons between our fitted V-band afterglow
and the observations 
not corrected for  extinction with the host galaxy's contribution 
subtracted, for GRB 011121 at $\rm z=0.36$. 
The CB's AG (the line of squares) is given by Eqs.~3-5~.  
The contribution
from a 1998bw-like supernova placed at the GRB's
redshift and modified by Galactic extinction, Eq.~(\ref{bw}), 
is indicated by a line of crosses.}
\label{figv1121}
\end{figure}  

%%%%%%%%%%%%%%%%%%%%%%%%%%%%%%%%%%%%%%%%%%%%%%%%%%%%%%%%
\begin{figure}[]
\hskip 2truecm
\vspace*{2cm}
\hspace*{-1.7cm}
\epsfig{file=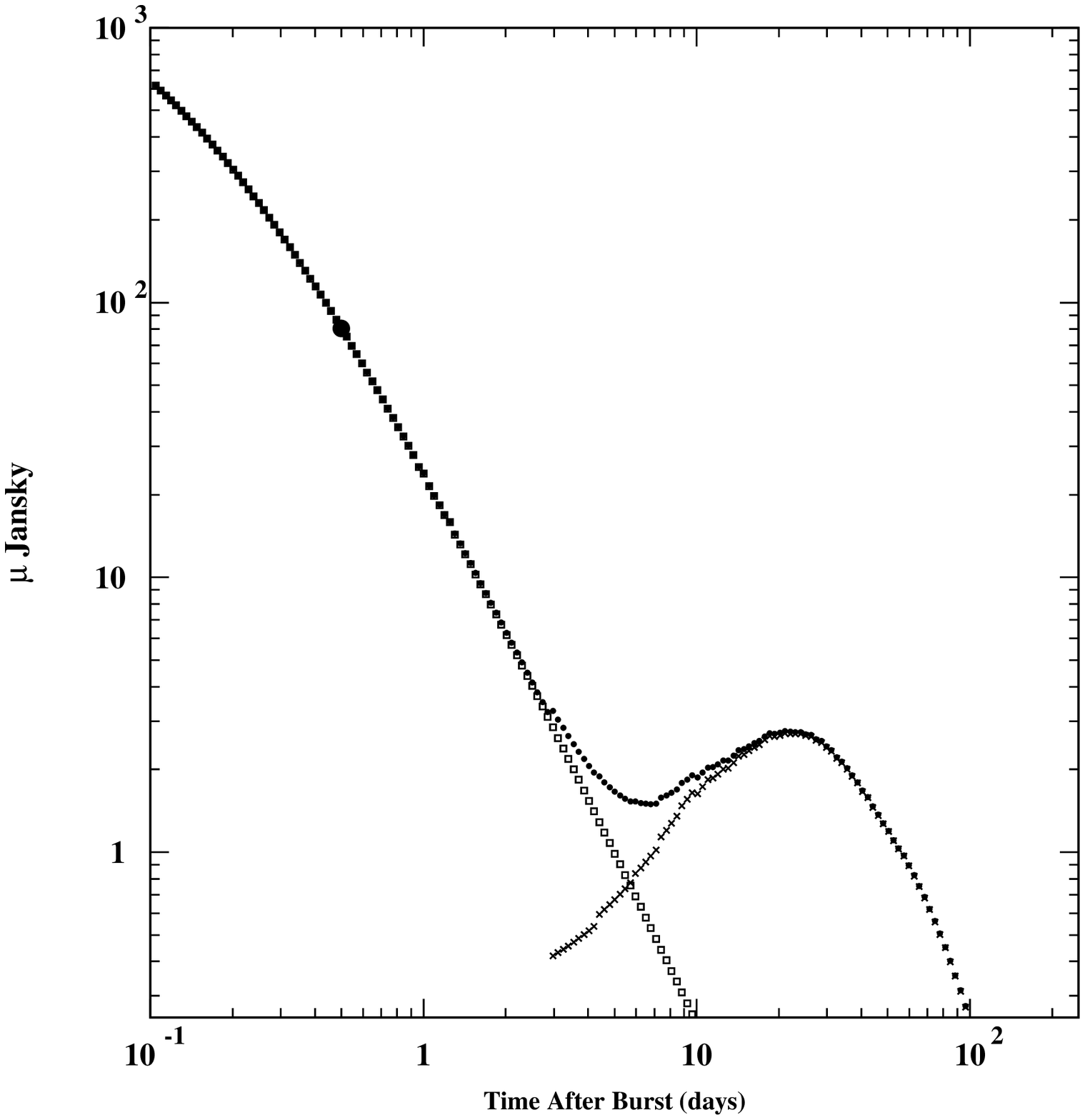, width=16cm} 
\caption{ Comparisons between our fitted I-band afterglow
and the observations 
not corrected for  extinction with the host galaxy's contribution 
subtracted, for GRB 011121 at $\rm z=0.36$. 
The CB's AG (the line of squares) is given by Eqs.~3-5~.  
The contribution
from a 1998bw-like supernova placed at the GRB's
redshift and modified by Galactic extinction, Eq.~(\ref{bw}), 
is indicated by a line of crosses.}
\label{figi1121}
\end{figure}  

\end{document}